\documentclass[
reprint,
superscriptaddress,
footinbib,
amsmath,amssymb,
aps,
prb
]{revtex4-1}
\usepackage{graphicx}
\usepackage{array}
\usepackage{dcolumn}
\usepackage{bm}
\usepackage[utf8]{inputenc}
\usepackage[english]{babel}
\usepackage{amsmath}    
\usepackage{graphicx}   
\usepackage{verbatim}   
\usepackage[usenames,dvipsnames]{color}      
\usepackage{subfigure}  
\usepackage{hyperref}   
\hypersetup{colorlinks=true,citecolor=blue,linkcolor=red,urlcolor=blue}
\usepackage{color}

\definecolor{green}{rgb}{0,0.5,0}

\usepackage{url}

\begin{document}

\title{Friction of Physisorbed Nanotubes: Rolling or Sliding?}

\author{\firstname{Davide} \surname{Mandelli}}
\affiliation{Atomistic Simulations, Italian Institute of Technology, via Morego 30, Genova 16163, Italy}

\author{\firstname{Roberto} \surname{Guerra}}
\email{roberto.guerra@unimi.it}
\affiliation{Center for Complexity and Biosystems, Department of Physics, University of Milan, via Celoria 16, 20133 Milano, Italy}


\begin{abstract}
The structure and motion of carbon and \textit{h}-BN nanotubes (NTs) deposited on graphene is inquired theoretically by simulations based on state-of-the-art interatomic force fields. Results show that any typical cylinder-over-surface approximation is essentially inaccurate. NTs tend to flatten at the interface with the substrate and upon driving they can either roll or slide depending on their size and on their relative orientation with the substrate. In the epitaxially aligned orientation we find that rolling is always the main mechanism of motion, producing a kinetic friction linearly growing with the number of walls, in turn causing an unprecedented supra-linear scaling with the contact area. A 30 degrees misalignment raises superlubric effects, making sliding favorable against rolling. The resulting rolling-to-sliding transition in misaligned NTs is explained in terms of the faceting appearing in large multi-wall tubes, which is responsible for the increased rotational stiffness. Modifying the geometrical conditions provides an additional means of drastically tailoring the frictional properties  in this unique tribological system. \href{https://dx.doi.org/10.1039/D0NR01016B}{DOI: 10.1039/D0NR01016B}
\end{abstract}

\maketitle

\section{Introduction}\label{sec.intro}

Nanotubes (NTs), either made of carbon or hexagonal boron-nitride (\textit{h}-BN), have been investigated with enormous interest in the last few decades due to their extraordinary mechanical and electronic properties. Nowadays, almost defectless NTs can be formed with lengths of 1\,cm or more \cite{Zhang2013}, and precise measurements of their mechanical and frictional properties have started to appear in literature \cite{Garel2012,Nigues2014}, revealing a rich variety of unexpected phenomena.
While the empirical laws of macroscopic friction are well known, the fundamental understanding of the tribological mechanisms at the microscopic scale is still under investigation from many points of view. The basic difficulty of friction is intrinsic, concerning the dissipative dynamics of systems with a large number of atoms, often involving ill-characterized sliding interfaces, which are buried and thus experimentally hardly accessible.

In a pioneering experiment by Falvo et al.\ \cite{Falvo1999}, both sliding and rolling motion of multi-walled (MW) carbon NTs on graphite were observed as a result of side pushing. Estimate of adhesion and of contact width based on cylinder-on-flat models led to the explanation that lattice registry between the NT contact region and graphite was the cause of rolling in some cases.
Due to the severe difficulties in manipulating such small objects, other investigations on these systems were mainly based on atomistic simulations \cite{Buldum1999,Chen2013,Li2014,Oz2016,Wang2019}, where to simplify the calculations the NT was often modeled by a single-wall (SW) treated as a rigid cylinder.
However, in recent years, new evidences have recognized that SWNTs do not maintain cylindrical shape when adsorbed on a surface, but rather tend to collapse in a ``dogbone'' like shape \cite{Xu2018,Impellizzeri2019}.
In the case of double-walled NTs (DWNTs) and of MWNTs in general, it has been recently shown that the interaction among the walls can be a source of spontaneous formation of facets \cite{Leven2016}. The effect of such faceting, strongly manifesting in large MWNTs (diameter $D>\sim10$\,nm) with small or no relative chiral angles between the NT walls \cite{Leven2016}, has been reported of critical impact on their frictional and mechanical properties \cite{Guerra2017}. Specifically, for NTs on flat surfaces, faceting can produce a fivefold increase of the contact area, thus enhancing the adhesive forces of the same proportion.

All the above factors have a considerable influence on the dynamics of NTs laterally pushed over a flat surface, ultimately determining whether they will slide or roll. In principle, one would expect rolling motion to be hindered by the presence of facets. Hence, since only the smallest MWNTs tend to keep a cylindrical shape, one might anticipate a rolling-to-sliding transition around the threshold size for faceting formation. On the other hand, lateral friction forces generally increase with the contact area, thus making rolling favorable in the macroscopic limit. Furthermore, changing the relative lattice orientation of nanoscale contacts can lead to gigantic effects on their dynamical response to external forces \cite{Dienwiebel2004}. These considerations suggest that the apparently simple rolling-or-sliding question hides a complex interplay between many phenomena, each of them subtly depending on the NT/substrate contact geometry. Accurate simulations able to account for the NT internal degrees of freedom and of many-body terms in the interatomic interactions are therefore requested to shed light on the tribological mechanisms taking place in these systems.

In this work we make use of molecular dynamics simulations to inquire theoretically the motion of carbon and \textit{h}-BN DWNTs and MWNTs laterally pushed over a graphene plane.
For \textit{h}-BN on graphene, the inherent lattice mismatch between the two materials enables us to investigate different contact configurations, which unravel a rich dynamical behavior. In particular, we anticipate that by controlling the NT size and their angular alignment relative to the substrate crystalline axis, one can induce different types of motion: from pure rolling in aligned geometries, to pure sliding in faceted, misaligned MWNTs, to a mixed rolling+sliding in collapsed misaligned DWNTs. As discussed in the next, the resulting motion can be understood in terms of contact geometry and energetics.

Results for carbon nanotubes are provided and discussed for comparison in Section \ref{sec.carbon}.

\begin{figure}[tb]
  \centering
  \includegraphics[width=\columnwidth]{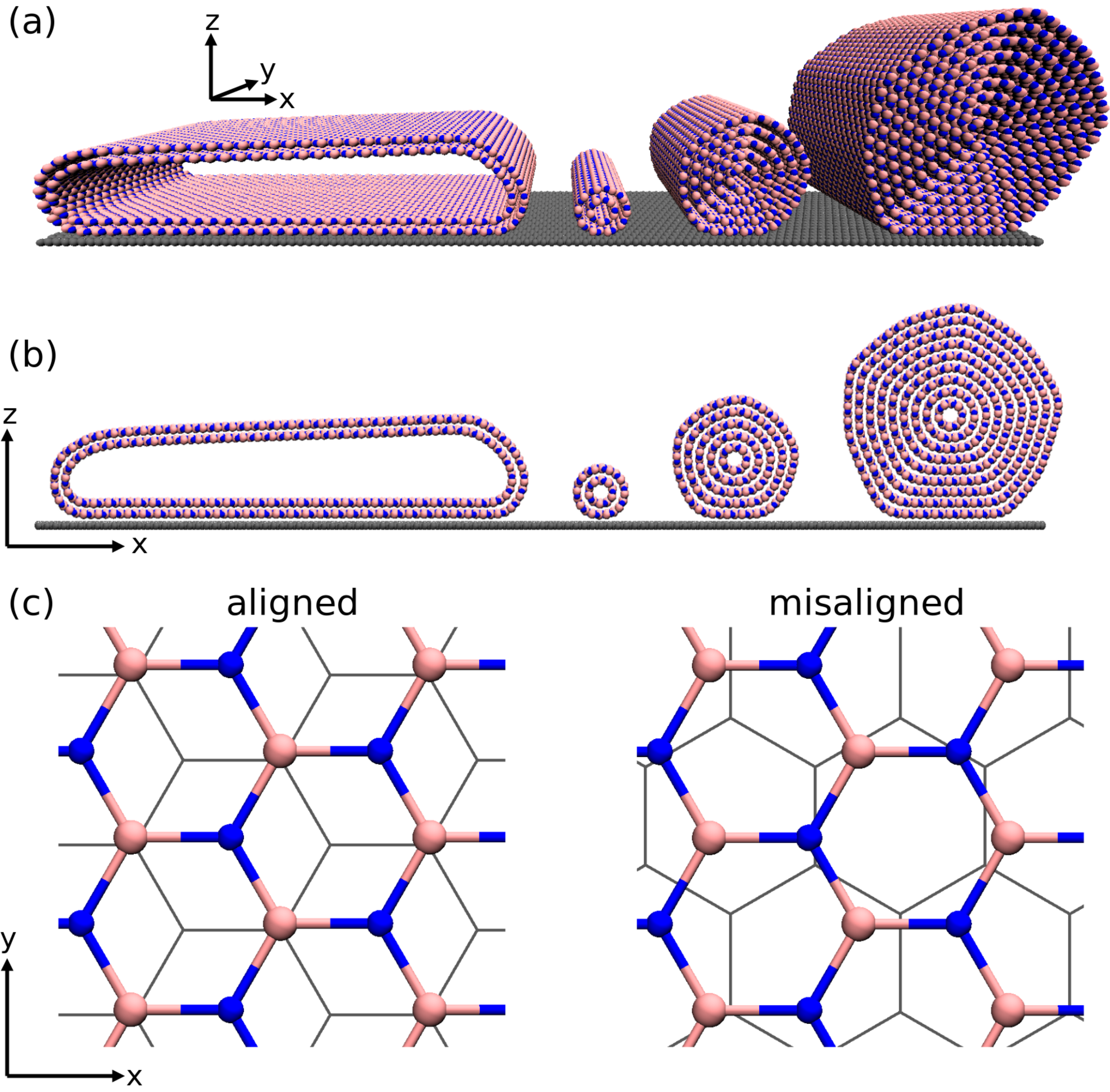}
  \caption{\small (a) Perspective view and (b) cross section of some of the double-walled and multi-walled \textit{h}-BN nanotubes
                  physisorbed on graphene used in simulations. From left to right, we report the relaxed structures
                  of the $65@70$ DWNT and of the $5@10$, $5@10...@25$ and $5@10...@45$ MWNTs in the $30^\circ$ misaligned
                  contact geometry. Carbon, boron and nitrogen atoms are colored in gray, pink and blue, respectively.
                  Panel (c) depicts a portion of NT/substrate interface in the aligned and $30^\circ$ misaligned configurations.}
  \label{fig1.schematics}
\end{figure}

\section{System and Methods}

{\bf Numerical modeling} --
The model systems studied consist of armchair \textit{h}-BN nanotubes of different sizes physisorbed on a rigid graphene layer (see figure~\ref{fig1.schematics}(a),(b)). In our reference coordinates frame the graphene substrate lies in the ($x$,$y$) plane, while the NT axis is along $y$, parallel to the zigzag axis.
We consider two different contact geometries. In ``aligned'' contacts, the zigzag direction of the graphene substrate is also
oriented along $y$, while in the $30^\circ$ ``misaligned'' contacts, it is oriented along $x$ (see figure~\ref{fig1.schematics}(c)).
Periodic boundary conditions are applied in $x$,$y$ directions, using rectangular supercells of suitably chosen sides ($L_x$,$L_y$).
Consequently, any rotation of the NT axis (misalignment) is hindered. This choice was made in order to better study the main frictional response in the opposite ideal cases of locally commensurate and incommensurate interfaces.
Also, periodicity conveniently removes any edge effect, which can play some role in short NTs, not discussed here.

NTs are built with a boron-nitrogen bond length of $d_{\rm BN}=1.442$\,\AA\ (honeycomb lattice constant $a_{\rm BN}=\sqrt{3}d_{\rm BN}\approx2.498$\,\AA), corresponding to the equilibrium value of the adopted Tersoff potential \cite{Sevik2011}.
The lattice constant, $a_{\rm g}$, of the rigid graphene substrate is chosen so to closely reproduce the experimental lattice mismatch $a_{\rm BN}/a_{\rm g}\simeq1.018$  \cite{Woods2014}, while also allowing the matching of NT and graphene supercells along the NT axis direction $y$.

In aligned contacts, the relevant zigzag $y$-periodicities of the substrate and of the NT are equal to $a_{\rm g}$ and $a_{\rm BN}$, respectively (see figure~\ref{fig1.schematics}(c), left panel). In this case, we fix $a_{\rm g}=\frac{54}{55}a_{\rm BN}\simeq2.452$\,\AA\ and we construct supercells of side $L_y=55a_{\rm g}=54a_{\rm BN}\simeq134.87$\,\AA.
In misaligned contacts, the substrate $y$-periodicity changes to the armchair value of $\sqrt{3}a_{\rm g}$ (see figure\ref{fig1.schematics}(c), right panel). In this case, we set $a_{\rm g}=2.4514$\,\AA, which allows constructing commensurate supercells of side $L_y=10\sqrt{3}a_{\rm g}=17a_{\rm BN}\simeq42.46$\,\AA.
We note that the adopted lattice constants are close to the corresponding experimental values, $a_{\rm g}^{\rm exp}\approx2.46$\,\AA\, and $a_{\rm BN}^{\rm exp}\approx2.50$\,\AA\ \cite{Lynch1966}.
For all the considered systems we use a graphene supercell side $L_x\simeq 300$\,\AA, which is large enough to avoid self-interaction of NTs with their periodic replica.
Tables~\ref{tab1.geoDW} and \ref{tab2.geoMW} report the geometric parameters of the DW and MWNTs considered.
Chiral indexes of BNNTs have been chosen so to obtain inter-wall distances close to the theoretical equilibrium bulk distance
of $\simeq3.34$\,\AA\ of the adopted inter-layer potential~\cite{Leven2014a,Leven2016c}. The reported nominal diameters $D$ are for the cylindrical configurations before relaxation.

For brevity, in the text we label the NTs indicating a single chiral index for each armchair wall, e.g., $15@20\equiv(15,15)@(20,20)$.
Following the discussion, the smallest $5@10$ DWNT appears in the list of MWNTs.
\begin{table}
\centering
\begin{tabular}{ c  c  c }
\hline
\hline
 $n_1$ & $n_2$ & $D$ (\AA) \\
\hline
 15 & 20 & 27.5402 \\
 25 & 30 & 41.3103 \\
 35 & 40 & 55.0803 \\
 45 & 50 & 68.8504 \\
 55 & 60 & 82.6205 \\
 65 & 70 & 96.3906 \\
\hline
\end{tabular}
\caption{\small The chiral indices $n_1$, $n_2$, and the nominal diameter $D$ of the simulated $(n_1,n_1)@(n_2,n_2)$ armchair \textit{h}-BN DWNTs.}
\label{tab1.geoDW}
\end{table}
\begin{table}
\centering
\begin{tabular}{ c  c  c }
\hline
\hline
 $N_{\rm w}$ & $n_{\rm out}$ & $D$ (\AA) \\
\hline
 2 & 10 & 6.88500 \\
 3 & 15 & 10.3276 \\
 4 & 20 & 13.7701 \\
 5 & 25 & 34.4252 \\
 6 & 30 & 41.3103 \\
 7 & 35 & 48.1953 \\
 8 & 40 & 55.0803 \\
 9 & 45 & 61.9654 \\
\hline
\end{tabular}
\caption{\small The total number $N_{\rm w}$ of walls, the chiral index $n_{\rm out}$ of the outermost wall, and the nominal diameter $D$ of the simulated $(5,5)@(10,10)@(15,15)...@(n_{\rm out},n_{\rm out})$ armchair \textit{h}-BN MWNTs.}
\label{tab2.geoMW}
\end{table}

{\bf Simulation protocol} --
Boron-nitrogen intra-layer interactions are modeled by a Tersoff potential as parametrized in Ref.~27.
Inter-layer interactions among the NT walls and between NT walls and graphene are modeled via the many-body force-field described in Refs.~30,31.
Since interaction between second-neighbor layers is negligible \cite{Leven2016c}, we reduce the computational burden by considering only the interaction among nearest-walls. Similarly, we only consider the interaction of graphene with the outermost wall of the physisorbed NT.

In our simulations all the NT atoms are free to move in every direction under the action of the external driving force and of the internal inter-atomic interactions. Hence, dynamical deformations are fully taken into account.

Fully relaxed configurations are obtained via geometry optimizations by numerically propagating the equation of motion using the standard velocity-Verlet integrator (time step $\Delta t=1$\,fs), coupled to the FIRE energy minimization algorithm \cite{Bitzek2006}. Simulations are stopped when the absolute value of the forces acting on each atom are all below the threshold value of  $10^{-3}$\,meV/atom.

For each NT configuration, we compute the corresponding sliding potential energy traces. Following a quasi-static protocol, we perform a sequence of simulations where the NT structure relaxed at previous step is rigidly displaced along $x$ by a small amount, $\Delta x=0.1062$ and $0.064$ \AA\ for aligned and misaligned configurations, respectively.
All atomic positions are subsequently relaxed while fixing the $x$ position of the NT center of mass (c.o.m.). The procedure is repeated until at least one full period of the energy trace is obtained.
Note that since in our setup energetics depends very poorly on $y$ coordinate -- the system having a long range commensuration along that direction -- in the following we report energy and force traces only pertaining to the motion direction $x$.

Dynamical simulations are performed propagating the Langevin equation of motion
\begin{equation}
\label{eqnew}
m_i \ddot {\bf r}_i={\bf F}_i-m_i\gamma({\bf v}_i-{\bf v}_{\rm cm})+F_{\rm ext}\hat x
\end{equation}
where ${\bf r}_i$ is the position of the $i$-th atom, ${\bf v}_i$ and $m_i$ are its velocity and mass, ${\bf F}_i$ is the total force due to the chosen set of interatomic potentials and ${\bf v}_{\rm cm}$ is the NT c.o.m.\ velocity. The second term in the r.h.s.\ is a viscous drag ($\gamma=0.1$\,ps$^{-1}$) used to avoid system overheating. Following Ref.~13, 
the dynamic friction force is evaluated from the shear force required to keep the NT at constant velocity $v_{\rm ext}=1$\,m/s along $x$ direction. To this end, we apply an external uniform force
\begin{equation}
\label{eq3}
F_{\rm ext}=\frac{\bar m}{\Delta t}\left(v_{\rm ext}-{\bar v}\right)+{\bar m}\gamma\left({\bar v}-v_{{\rm cm},x}\right)-{\bar m}{\bar a}
\end{equation}
to each of the $N$ atoms of the outermost wall of the NT, where ${\bar m}=N\left(\sum_{i=1}^N\frac{1}{m_i}\right)^{-1}$, ${\bar a}=\frac1N\sum_{i=1}^N\frac{F_{i,x}}{m_i}$, ${\bar v}=\frac1N\sum_{i=1}^Nv_{i,x}$ and $\Delta t=1$\,fs is the numerical propagation time step. Note that after the first time step, $v_{\rm ext}={\bar v}\approx v_{{\rm cm},x}$, so that $F_{\rm ext}\approx-{\bar m}{\bar a}$, i.e.\,, minus the total force per atom acting on the driven wall.

Since the viscous damping is applied to all the NT atoms but not to its c.o.m.\ motion, the computed friction results weakly dependent on the adopted $\gamma$ value, the latter mainly determining the steady-state temperature of the sliding system. In our typical simulations, which are run in the underdamped regime, we measure steady-state temperatures well below $0.01$\,K, indicating a negligible role of thermal excitations on the measured friction.

To check the sensitivity of the results on the driving protocol, we have performed additional simulations in which the NT is pushed by a repulsive wall or by a repulsive tip moving at constant velocity. Since the different driving protocols produced equivalent results, in the following we report only data obtained with a uniform external force, equation~\ref{eq3}.

{\bf Definitions} --
Since $F_{\rm ext}$ is applied uniformly to all the $N$ atoms of the outermost tube, the instantaneous total friction force, $F_{\rm fric}$, is simply expressed by
\begin{equation}\label{eq7}
F_{\rm fric}=NF_{\rm ext}.
\end{equation}
The kinetic friction force, $F_{\rm kinetic}$, is obtained by averaging $F_{\rm fric}$ during steady-state motion, after the initial transient dynamics has decayed, over a time window covering an integer number of oscillations of the periodic force traces.

Instead, the static friction force, $F_{\rm static}$, is extracted from the peak value of the friction force trace during steady-state.
Equivalent results are obtained estimating $F_{\rm static}$ from the maximum derivative of the potential energy traces.

We estimate the contact area $A$ of the fully relaxed NT/graphene interface using as reference the adhesion energy per unit area calculated for an infinite \textit{h}-BN/graphene plane bilayer, $\varepsilon_{\rm adh} = 1.812$\,eV/nm$^2$ and $1.786$\,eV/nm$^2$ for aligned and misaligned configurations, respectively:
\begin{equation}\label{eq8}
A=\frac{E_{\rm adh}}{\varepsilon_{\rm adh}}~,
\end{equation}
where the adhesion energy of the NT/graphene system, $E_{\rm adh}=E^{\rm NT/g}-E^{\rm NT}-E^{\rm g}$, is the difference between the total energy, $E^{\rm NT/g}$, of the relaxed NT/graphene system and the total energies $E^{\rm NT}$, $E^{\rm g}$ of the separate NT and substrate. Note that quantitatively similar $A$ values are obtained by defining the NT contact atoms as those with a distance smaller than $4$\,\AA\ from the substrate, which we take as the reference threshold distance for having contact.

To describe quantitatively the rolling/sliding motion of the NT, we compute the average velocity along the $x$ sliding direction
\begin{equation}
\langle v_{\rm contact}\rangle = \frac{1}{N_{\rm contact}}\sum_{i=1}^{N_{\rm contact}}v_{i,x}~,
\end{equation}
considering the $N_{\rm contact}$ atoms of the NT belonging to the contact region defined above. The average is taken at steady-state, over a time window covering an integer number of oscillations of the periodic force traces. By normalizing with respect to the imposed sliding velocity, $v_{\rm ext}$, we obtain a scalar quantity, $\eta=\langle v_{\rm contact}\rangle/v_{\rm ext}\in[0,1]$.
A value of $\eta=0$ is obtained when the NT atoms at the interface do not move ($\langle v_{\rm contact}\rangle = 0$), i.e.\ pure rolling motion. On the other hand, $\eta=1$ corresponds to NT contact atoms moving at full speed, i.e.\ pure sliding. Intermediate values of $0<\eta<1$ correspond to a mixed rolling and sliding motion. Direct inspection of the trajectories supports this classification, validating the use of $\eta$.
Detailed insights on the internal NT dynamics are obtained by monitoring the angular velocity $\omega$ of each wall.
Finally, we define the relevant $x$-periodicity of the graphene substrate in the sliding direction $a_{\rm sub}=\sqrt{3}a_g$ and $a_{\rm sub}=a_g$, respectively for aligned and misaligned geometries (see figure~\ref{fig1.schematics}(c)).

\begin{figure}[tb]
  \centering
  \includegraphics[width=\columnwidth]{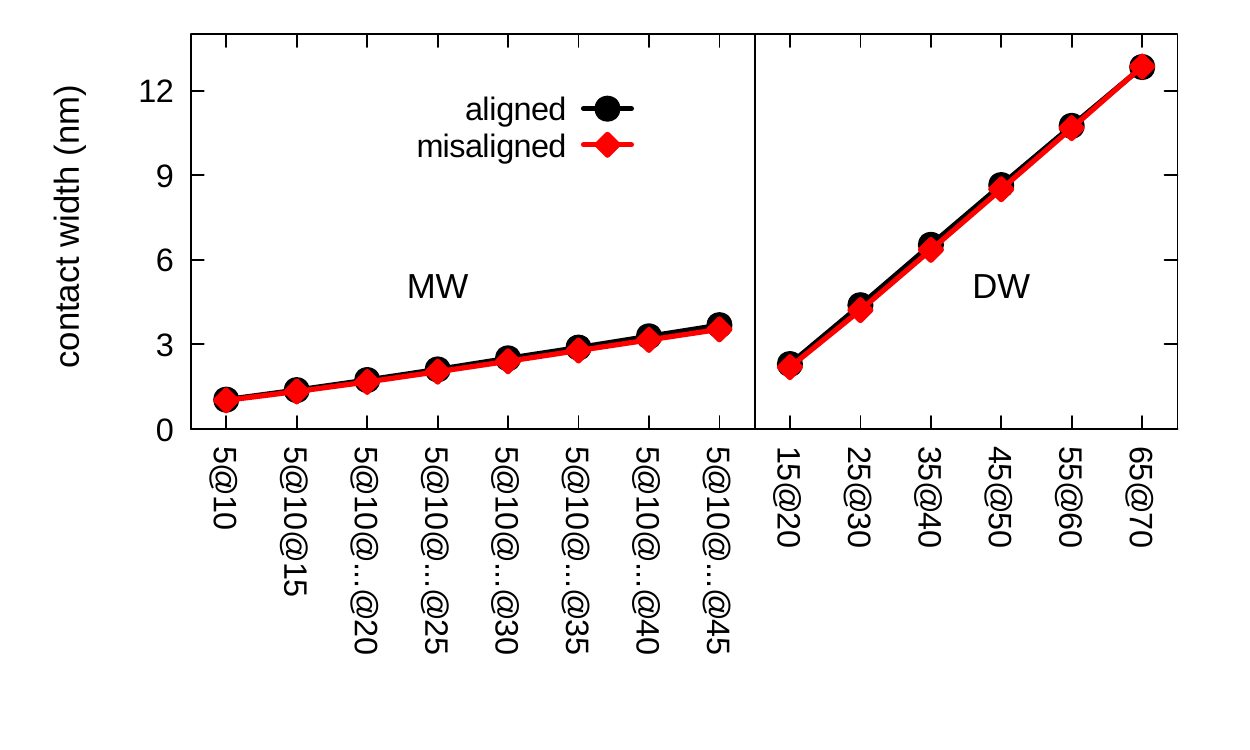}
  \caption{\small Contact width of the fully relaxed, aligned (black) and 30$^\circ$ misaligned (red) NTs.}
  \label{fig.width}
\end{figure}

\section{Results}

\subsection{Static properties and energetics}\label{sec.static}

Upon relaxation, large-enough MWNTs form a longitudinal faceting (see figure~\ref{fig1.schematics}(a),(b)) that allows energetically favourable stacking between neighboring walls while gathering most of the curvature stress in localized corners, between one face and the next. As predicted in previous work~\cite{Leven2016}, we found that the number of facets is dictated by the difference in the unit cells between mating layers, five in the present case. Differently, the more flexible DWNTs tend to collapse so to maximize adhesion energy, although no full junction between the opposite parts of the internal wall occurs, which in principle could lower total energy even more.

In all NTs considered, the contact area $A$ and the associated contact width $W=A/L_y$ grow linearly with the nominal diameter $D$, as reported in figure~\ref{fig.width}.
This is a direct geometrical consequence of the faceting of MWNTs and of the full collapse of DWNTs.
Furthermore, we found that the contact width is in practice independent of the alignment of the NT axis relative to the crystallographic directions of the substrate. This result reflects the weak angular dependence of the adhesion energy in twisted graphene/\textit{h}-BN heterojunctions~\cite{Mandelli2019}.
\begin{figure}[tb]
  \centering
  \includegraphics[width=0.915\columnwidth]{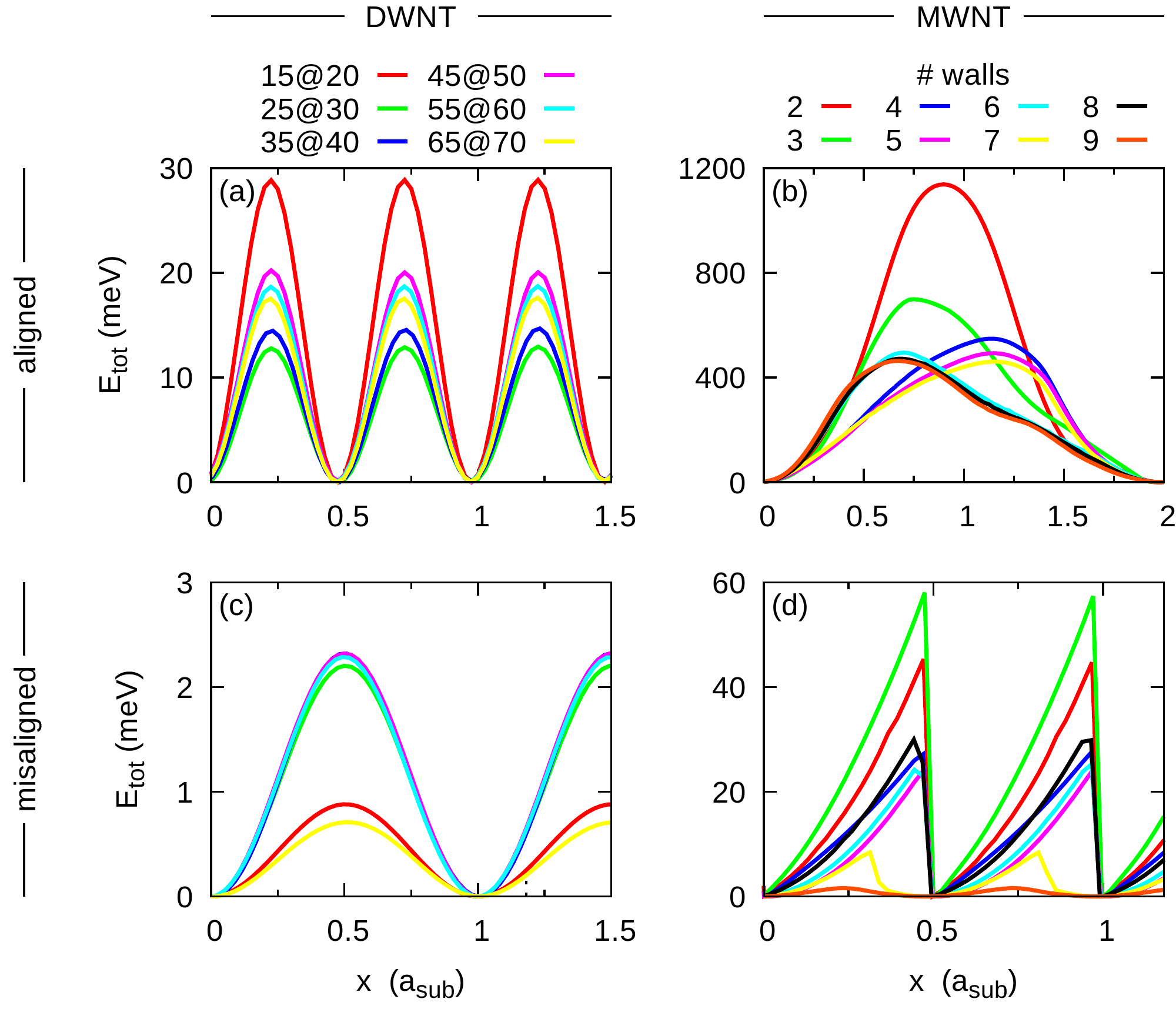}
  \caption{\small Potential energy profile obtained from a quasi-static displacement of (a),(c) DWNTs and (b),(d) MWNTs for (a),(b) aligned and (c),(d) misaligned configurations.}
  \label{fig.pes}
\end{figure}
\begin{figure}[tb]
  \centering
  \includegraphics[width=\columnwidth]{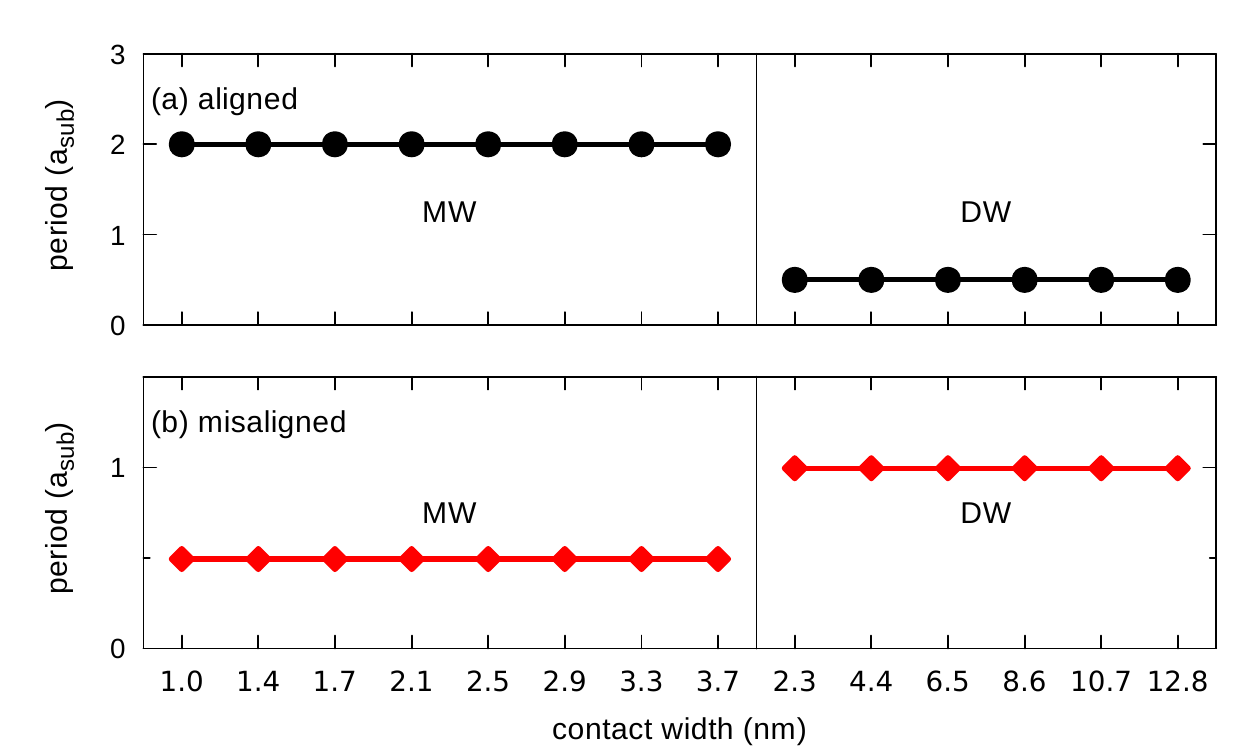}
  \caption{\small The periodicity of the adiabatic potential energy traces calculated for the aligned (a) and misaligned (b) configurations.}
  \label{fig.trace-periodicity}
\end{figure}

Before investigating the dynamical response to an external force, it is interesting to consider the pure energetics obtained by an adiabatic displacement of the NTs (see Methods). The resulting trajectories provide information about the type of motion in the zero-velocity limit, while the associated energy traces carry information such as static friction and periodicity. In figure~\ref{fig.pes} we report the calculated energy traces for the considered set of NTs. We note that DWNTs are characterized by a smooth variation of potential energy upon displacement (see figure~\ref{fig.pes}(a),(c)), following a sinusoidal-like trend, with periodicity depending exclusively on the alignment condition (see figure~\ref{fig.trace-periodicity}). The adiabatic trajectories show that DWNTs advance by rolling smoothly over the substrate (see Supplementary Movies 1,2 $\dag$). In the case of MWNTs, the different alignment not only affects the periodicity (see figure~\ref{fig.trace-periodicity}), but also results in a completely different shape of the energy traces (see figure~\ref{fig.pes}(b),(d)): while aligned MWNTs show smooth (although irregular) trends, misaligned MWNTs are characterized by a steady rise of the energy followed by a sharp drop, suggesting a different type of motion. Inspection of the adiabatic trajectories reveals pure rolling of aligned MWNTs and mixed rolling+sliding motion accompanied by sudden backward rotational slip events of misaligned MWNTs (see Supplementary Movies 3,4,5,6 $\dag$).
\begin{figure}[tb]
  \centering
  \includegraphics[width=0.915\columnwidth]{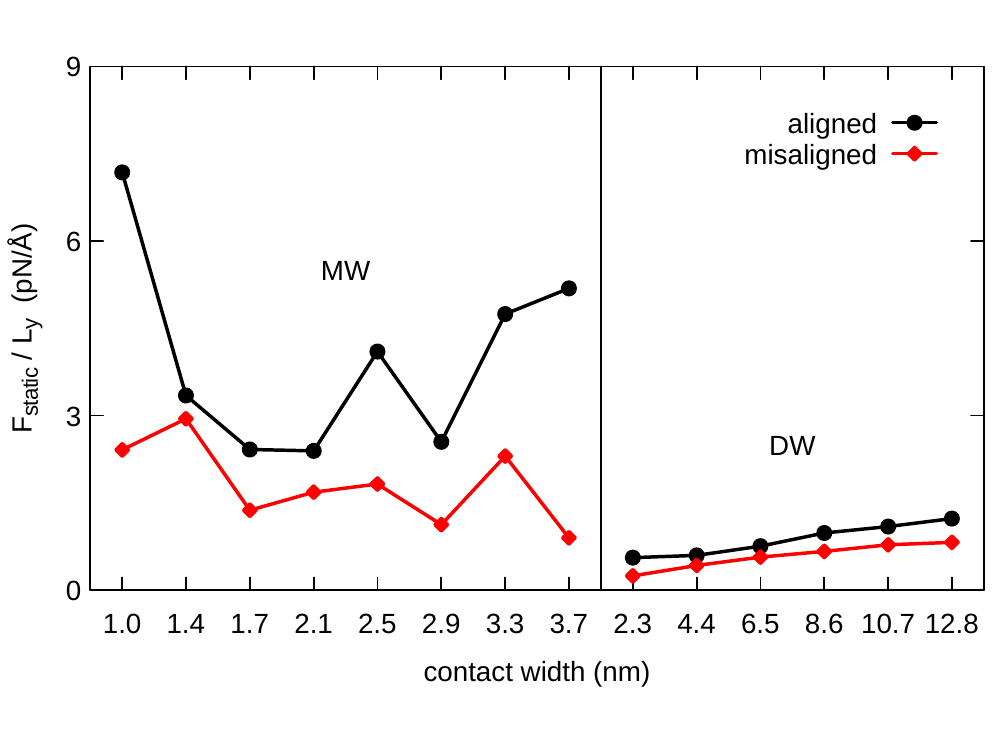}
  \caption{\small Static friction force per unit length of the NTs in the aligned (black) and misaligned (red) configurations. The lines are drawn to better distinguish the data sets.}
  \label{fig.friction_static}
\end{figure}
\begin{figure}[tb]
  \centering
  \includegraphics[width=0.915\columnwidth]{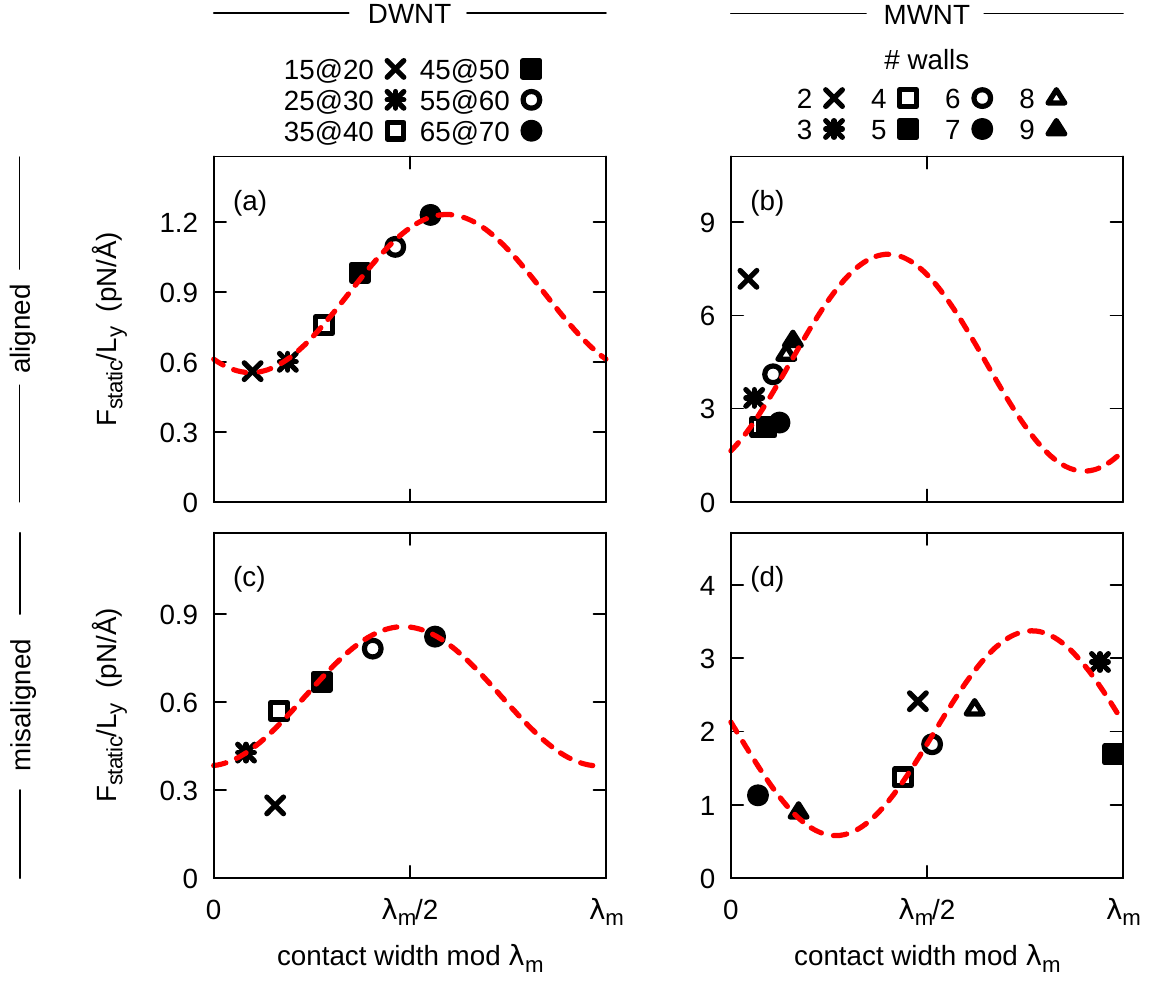}
  \caption{\small Static friction force per unit length of (a),(c) DWNTs and (b),(d) MWNTs for (a),(b) aligned and (c),(d) misaligned configurations. Data are reported as a function of the contact width $W$, modulo the period of the NT/graphene interfacial moir\'e pattern, $\lambda_{\rm m}$. Red dashed lines are fit to $F_{\rm static}(W)=\alpha+\beta\sin(\frac{2\pi}{\lambda_{\rm m}}W+\phi)$, from which the moir\'e periodicity has been estimated.}
  \label{fig.Fs-moire}
\end{figure}

It is worth to note that within each of the four considered groups (panels of figure~\ref{fig.pes}), the height of the barrier opposing to motion -- directly connected to the static friction force -- seems to vary irregularly with the NT size. In the conventional picture of friction one could expect instead a sublinear growing of static friction with the contact area \cite{DeWijn2012,Varini2015,Mandelli2018c}, and thus with $D$. As reported in figure~\ref{fig.friction_static}, in the present case we observe a static friction force per NT length, $F_{\rm static}/L_y$, of large magnitude and strongly fluctuating with $D$ for MWNTs, while of reduced intensity and of smoother trend for DWNTs.

This apparently erratic behavior can be rationalized in terms of the partial compensation of lateral forces occurring in finite size incommensurate contacts~\cite{Gigli2017}, as follows. Due to the inherent lattice mismatch between graphene and \textit{h}-BN, a characteristic moir\'e superstructure appears at the NT/graphene interface~\cite{Guerra2017b}, whose periodicity in the $x$ sliding direction is equal to
\begin{equation}\label{eq.lambda}
\lambda_{\rm m}=\sqrt{3}a_{\rm BN}/(\sqrt{3}a_{\rm BN}-a_{\rm sub}),
\end{equation}
where $\sqrt{3}a_{\rm BN}$ and $a_{\rm sub}$ are the relevant NT and substrate periodicities, respectively (see Methods).
As shown in recent work~\cite{Gigli2017}, for contacts of width $W$ close to multiple integers of $\lambda_{\rm m}$, good compensation of forces occurs that minimizes static friction. On the other hand, maximum pinning to the substrate is observed for values of $W$ that are close to half odd-integers of $\lambda_{\rm m}$. Overall, this gives rise to an oscillatory behavior of the static friction force that can be modeled as
\begin{equation}\label{eq.Fs-period}
F_{\rm static}=\alpha+\beta\sin\left(\frac{2\pi W}{\lambda_{\rm m}}+\phi\right).
\end{equation}
We note that this behavior has been demonstrated so far only for the static friction force opposing the sliding of flat contacts. However, as we will show in the following, our results suggest that equation~\ref{eq.Fs-period} is able to describe also the barriers against rolling.

In figure~\ref{fig.Fs-moire}(a), we report the contact width dependence of the static friction force extracted from the dynamical simulations of the aligned DWNTs, showing the expected sinusoidal trend. A fit of the data via equation~\ref{eq.Fs-period} yielded a value of $\lambda_{\rm m}\sim23$ nm, in agreement with the nominal value of $23.36$ nm predicted by equation~\ref{eq.lambda}. The same value of $\lambda_{\rm m}$ also provides a reasonable description of the static friction force measured in the aligned MWNTs (see figure~\ref{fig.Fs-moire}(b)), with the exception of the smallest $5@10$ NT (cross symbol). We note, however, that in the latter case the value of the contact width is not well defined due to the reduced size and cylindrical shape of the NT (see figure~\ref{fig1.schematics}).
The analysis of the type of motion (figure \ref{fig.pes}), of the periodicities (figure \ref{fig.trace-periodicity}), and of the friction level (figure \ref{fig.friction_static}), clearly indicates that the $5@10$ NT belongs to the class of MWNTs, rather than that of DWNTs. Such behavior is due to the retain of the cylindrical shape and the impossibility of collapsing as occurring in larger DWNTs.

Similar analysis for the $30^\circ$ misaligned NTs (see figure~\ref{fig.Fs-moire}(c),(d)) yielded an estimate of $\lambda_{\rm m}\sim0.68$ nm, close to the nominal value of $0.57$ nm. The discrepancy can be attributed to the smaller moir\'e periodicity and the resulting strong oscillatory behaviour of the sinusoid in equation~\ref{eq.Fs-period}, which makes the fit more sensitive on the precise values of the contact width. The latter has been estimated based on adhesion energy arguments considering the fully relaxed structures (see Methods). Dynamical deformations occurring during sliding affect the instantaneous contact width, contributing to the overall uncertainty.

Regarding the average static friction value, $\langle F_s \rangle = \alpha + \beta/2$, we note that the always-rolling DWNT show the smaller $\langle F_s \rangle$, while larger values are shown by MWNT, especially in the case of aligned MWNT in which rolling acts against faceting inducing energetically costly deformations. Instead, internal NT deformations are mostly absent in misaligned MWNT where the main motion occurs through sliding, yielding a reduced $\langle F_s \rangle$. Note that the sliding degree reported in figure~\ref{fig.avg-v_contact}, obtained from NT dynamics at finite velocity, as discussed in the next, is also representative of the so far discussed ``motion'' of the NTs in the adiabatic limit.

\subsection{Dynamical properties and friction}\label{sec.dynamics}
\begin{figure}[tb]
  \centering
  \includegraphics[width=\columnwidth]{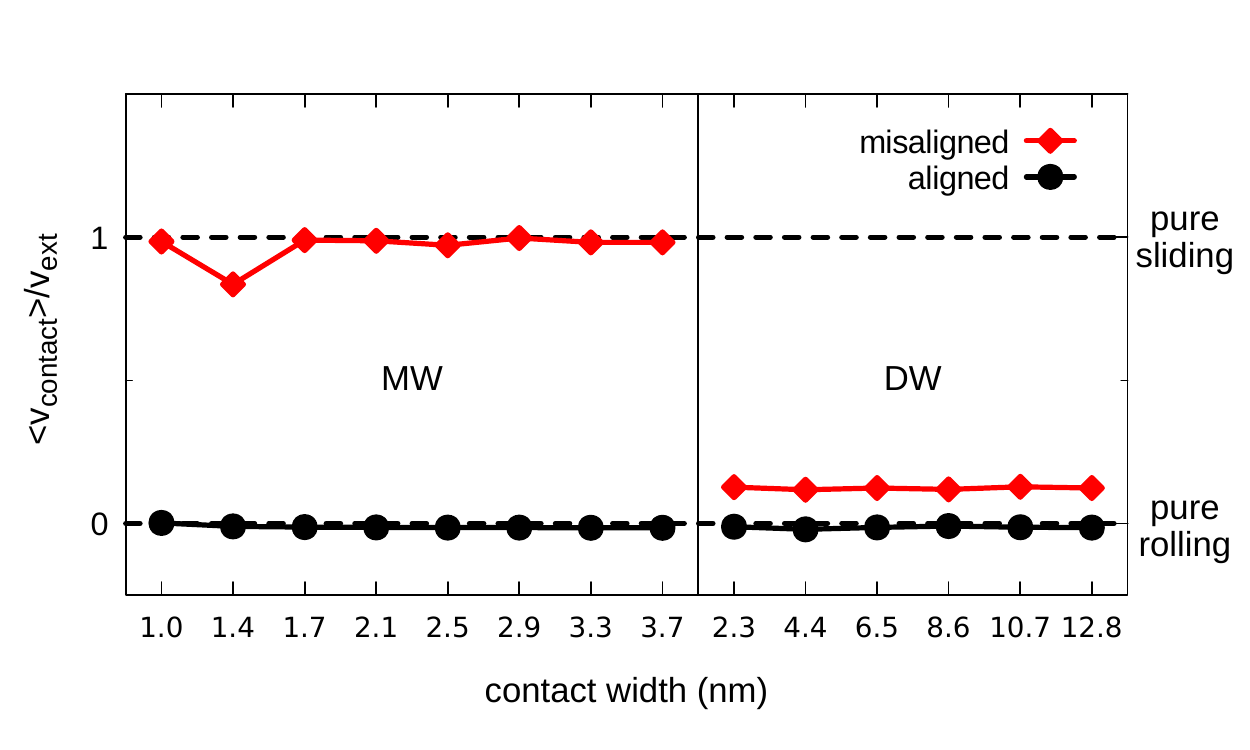}
  \caption{\small The normalized velocity of the contact atoms of the driven NTs in the aligned (black) and 30$^\circ$-rotated misaligned (red) configurations. Pure rolling and pure sliding limits are marked by dashed lines.}
  \label{fig.avg-v_contact}
\end{figure}

We now turn to the dynamics of NTs sliding at constant velocity atop the substrate. Upon driving, we observe that all the considered epitaxially aligned NTs show pure rotational motion, marked by a zero velocity of the contact atoms (see figure~\ref{fig.avg-v_contact}). We note that in all the cases rolling does not correspond to a rigid rotation of the relaxed NTs, but rather occurs through a belt-like motion of the walls so to preserve the same shape during advancement (see Supplementary Movies 1,3,5 $\dag$).
Instead, in misaligned NTs, contact atoms always present a non-zero sliding velocity component, which is maximized in the case of MWNTs (pure sliding), while becoming minimal for DWNTs (partial sliding). As anticipated above, this rolling to sliding transition can be explained in terms of the faceting appearing in large multi-wall tubes, which increases the rotational stiffness, thus favouring sliding.
No such transition is observed for epitaxially aligned NTs, due to a maximized corrugation energy which makes rolling much more convenient rather than sliding. As an example, the energy trace obtained via a rigid $x$-displacement (mimicking pure sliding) of the $5@10...@25$ aligned MWNT yielded a static friction force about tenfold larger than that obtained from adiabatic rolling.

Overall, in all systems investigated we observed similar kind of motions as predicted by the adiabatic trajectories described in previous section. This indicates that the dynamics of the systems investigated is mainly dictated by the energetics of the contact, inertia plays only a secondary role and determines fine features of the NT motion.

\begin{figure}[tb]
  \centering
  \includegraphics[width=\columnwidth]{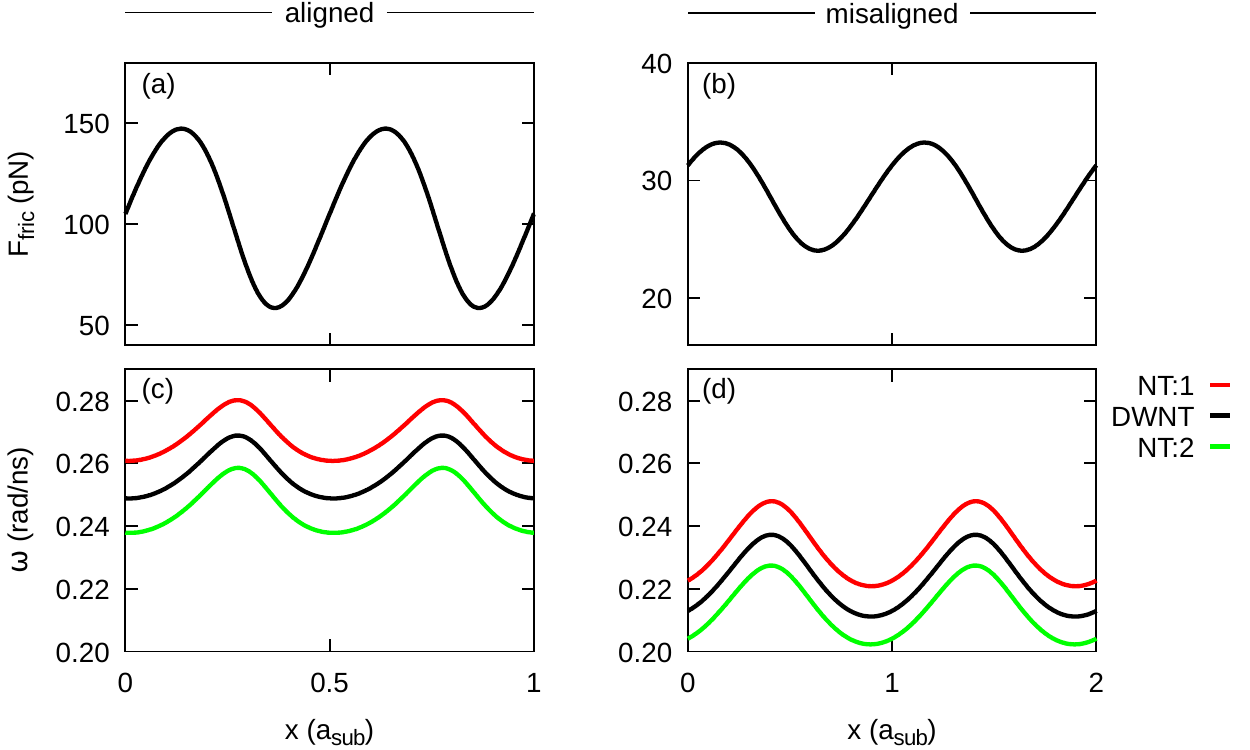}
\caption{\small DWNT -- Force traces and angular velocity of the individual walls (average in black) during lateral pulling of the $55@60$ DWNT. Left and right panels are for the aligned and misaligned configuration, respectively. Qualitatively similar results were obtained for all DWNTs considered.}
  \label{fig.traces.dwnt}
\end{figure}

\begin{figure}[tb]
  \centering
  \includegraphics[width=0.915\columnwidth]{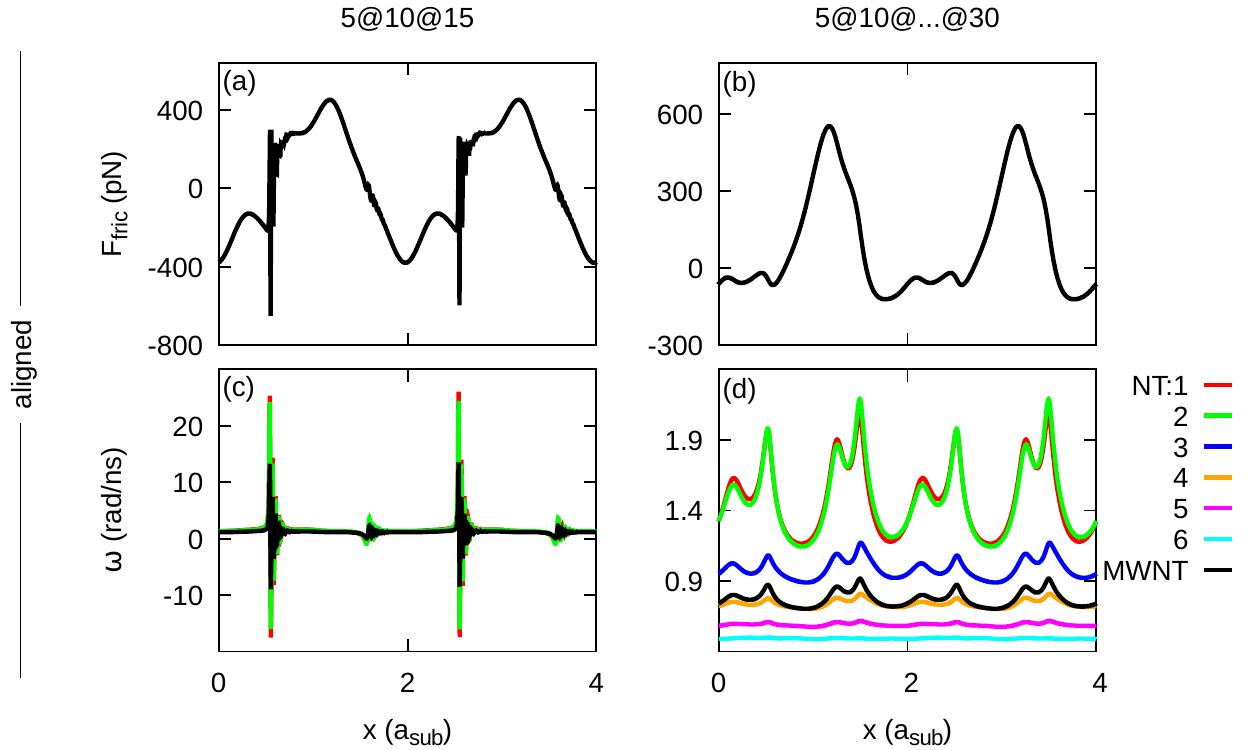}
  \includegraphics[width=0.915\columnwidth]{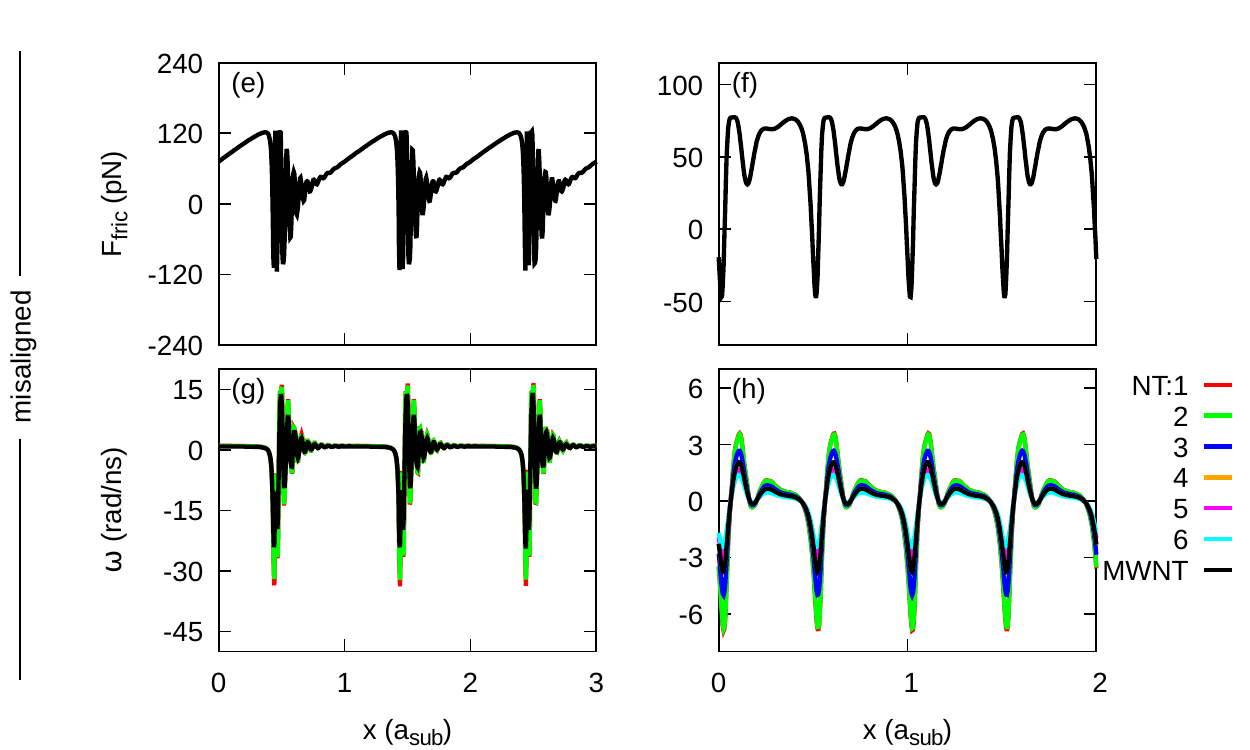}
  \caption{\small MWNT -- Force traces and angular velocity of the individual walls (average in black) during lateral pulling of (left panels) the $5@10@15$ and of (right panels) the $5@...@30$ MWNT. Panels (a)-(d) and (e)-(h) are for the aligned and misaligned configuration, respectively.}
  \label{fig.traces.mwnt}
\end{figure}

Detailed insights about the NT dynamics are gained by examining the dynamical friction traces $F_{\rm fric}$ and the angular velocities $\omega$ of each wall during sliding. These quantities are plotted in figures~\ref{fig.traces.dwnt} and  \ref{fig.traces.mwnt} for few representative DW and MWNTs, respectively. Considering at first the DWNTs, we observe that their smooth rolling motion is accompanied by a regular sinusoidal trend of both $F_{\rm fric}$ and $\omega$, as reported in figure~\ref{fig.traces.dwnt}. The residual sliding component present in the misaligned geometry (see figure~\ref{fig.avg-v_contact}) accounts for the reduction in the average angular velocity when compared to the aligned case (see black curves in figure~\ref{fig.traces.dwnt}(c),(d)). Notably, DWNTs do not roll as rigid bodies, the inner wall rotating faster than the outermost (see red and green curves in figure~\ref{fig.traces.dwnt}(c),(d)). This indicates an additional frictional contribution arising from inter-wall shear forces, which adds to the one due to the NT/substrate interactions.
In figure~\ref{fig.traces.mwnt}(a),(b) we report the friction traces of aligned MWNTs with 3 and 6 walls. Despite the analysis of the contact velocity indicating pure rolling of the outermost wall (see figure~\ref{fig.avg-v_contact}), the friction traces do not display a regular sinusoidal trend. The origin of this peculiar behavior is found in the faceting appearing in large NTs, which leads to periodic rearrangements of the inner walls during rolling. Specifically, for the case of the 5@10@15 NT, inspection of the trajectory reveals sudden backward rotations of the two inner walls. This is demonstrated by the negative peaks of the angular velocities reported in figure~\ref{fig.traces.mwnt}(c), which correlate with the abrupt changes of the friction force. These violent events disappear in systems of growing size, where the friction trace becomes smoother (see figure~\ref{fig.traces.mwnt}(b)). A clear indication of the inter-wall rearrangements occurring inside the faceted MWNTs is provided by the analysis of the angular velocities reported in figure~\ref{fig.traces.mwnt}(d), showing increasing angular velocity going from the outermost (cyan curve) to the innermost walls (red and green curves). As for the case of the DWNTs discussed above, this phenomenon represents an additional source of frictional dissipation.

A similar analysis for misaligned MWNTs is shown in figure~\ref{fig.traces.mwnt}(e)-(h). The friction trace of the small 5@10@15 NT (panel (e)) displays a saw-tooth behavior, where $F_{\rm fric}$ increases at a constant rate and subsequently drops. Inspection of the associated angular velocity shows that the rising phase corresponds to an initial forward rotation of the NT, followed, and only partially compensated, by a sudden backward rotational slip (panel (g), see also Supplementary Movie 4 $\dag$). This partial compensation is at the origin of the residual rolling component reported in figure~\ref{fig.avg-v_contact} for the misaligned 5@10@15 NT. In larger MWNTs characterized by marked faceting, the friction trace becomes more regular (see figure~\ref{fig.traces.mwnt}(f)) and the dynamics turns into a smooth sliding motion accompanied by a periodic back and forth rocking of the NT due to inertial effects (see angular velocities in panel (h) and Supplementary Movie 6 $\dag$).

\begin{figure}[tb]
  \centering
  \includegraphics[width=0.915\columnwidth]{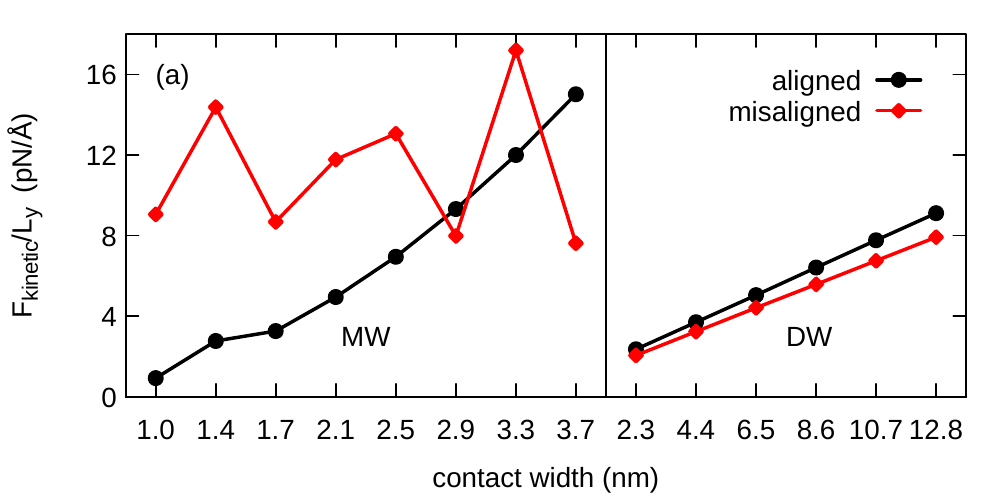}
  \includegraphics[width=0.915\columnwidth]{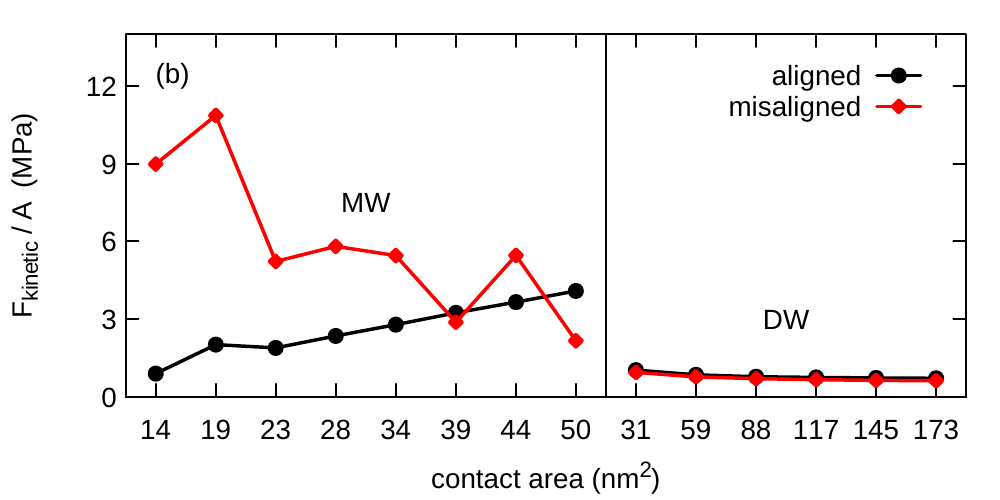}
  \caption{\small Total kinetic friction force (a) per unit length and (b) per unit area of the NTs in the aligned (black) and misaligned (red) configurations.}
  \label{fig.friction_dynamic}
\end{figure}

In figure~\ref{fig.friction_dynamic}(a) we plot the contact width dependence of the kinetic friction force per unit length of the considered set of NTs, showing different size scaling for the cases of rolling and sliding NTs. Considering first the case of rolling (aligned and misaligned DWNTs, aligned MWNTs), we can identify two main contributions to dissipation arising from NT/substrate interactions and NT internal deformations. Analysis of the static friction force -- mainly dictated by NT/substrate interactions -- demonstrated that the barrier against rolling follows an oscillatory behavior as a function of NT size, suggesting similar size independent contribution to kinetic friction. Internal deformations and inter-wall shear forces instead grow with NT diameter $D$, explaining the observed trend. Specifically, for the case of DWNTs, only one shearing interface is present, leading to linear dependence of $F_{\rm kinetic}/L_y$ on contact width $W\propto D$, and roughly constant kinetic friction stress, $F_{\rm kinetic}/A$ (see figure \ref{fig.friction_dynamic}(b)). On the other hand, in MWNT the number of shearing walls increases with NT size and internal deformations become more marked due to pronounced faceting. Overall, this leads to supra-linear increase of dissipation, as demonstrated in figure \ref{fig.friction_dynamic}(b), showing increasing kinetic friction stress with contact area $A$.

The case of misaligned MWNTs requires distinct discussion, because of the pure sliding occurring in this case. Indeed, it is known that the increasing contact width allows better sampling of the interface incommensurability, asymptotically leading to superlubricity in the thermodynamic limit of infinite interface size \cite{Peyrard2000,Wang2019b}. This means that for stiff, shearing incommensurate contacts, as those considered here, sliding friction must grow -- at most -- sublinearly with the contact area $A$, in agreement with figure~\ref{fig.friction_dynamic}(b) in the case of misaligned MWNTs. Besides, oscillations in the decreasing $F_{\rm kinetic}/A$ trend are likely due to the internal rearrangements of the individual walls taking place during motion, as discussed above.

\subsection{Carbon NT results}\label{sec.carbon}

Given the fact that homogeneous graphene/graphene and heterogeneous h-BN/graphene interfaces are characterized by quantitatively similar sliding potential and exfoliation energy landscapes \cite{Ouyang2018}, for the sake of comparison we have performed additional dynamical simulations of armchair carbon nanotubes (CNTs) on graphene in the 30$^\circ$ degrees misaligned geometry.
The epitaxially aligned case is less worth to investigate since CNT aligned on graphene form perfectly commensurate interfaces, thus leading to rolling always favourable against sliding, as already discussed above for BNNTs.

We have considered the 55@60 DWCNT and the 6-walls MWCNT as representative cases. The relaxation of these structures (see Supplementary Information Fig.~S1) produces a collapsing of the DWNT and a faceting of the MWNT comparable to those in Fig.~\ref{fig1.schematics}, although faceting being less pronounced in the carbon/carbon interface \cite{Leven2016}.
Dynamical simulations of the CNTs gave results equivalent to BNNT case. Namely, the DWCNT displays a belt-like motion with a small sliding component ($\eta \simeq 0.1$), while no rolling is observed in the MWCNT ($\eta = 1$). The friction force and angular velocity traces of each wall as a function of the CNT position during motion are also comparable (see Supplementary Information Fig.~S2). The MWCNT in particular shows a smoother dynamics when compared to the corresponding BNNT, likely due to a smaller interlayer corrugation energy. We further note that the traces of CNT and BNNT have the same periodicities (in units of the relevant substrate lattice constant).

We note that our case must be distinguished from that of telescopic sliding in which NT walls slide one relative to another \cite{Nigues2014}. There, friction was connected to a dynamical rearrangement of the facets \cite{Guerra2017}, more pronounced in large BNNTs due to their tendency of forming walls with similar chiral angles \cite{Leven2016}, and due to enhanced LA phonon scattering \cite{Prasad2017}.
No telescopic motion is expected in our case of NT sliding/rolling over a surface, in which the relative position among internal NT walls is mostly constant.

To conclude, our results for NTs over a flat surface suggest that CNT should behave similarly to BNNT, with a transition from almost pure rolling to pure sliding between misaligned DW and misaligned MWNT.

\section{Discussion}\label{sec.discussion}

We have simulated atomistically the induced motion of \textit{h}-BN DWNTs and MWNTs on graphene, considering the two ideally opposite cases of perfectly aligned and maximally misaligned interface crystal lattices.

We show that misaligned contacts systematically display a lower barrier against motion compared to the epitaxially aligned geometry, the latter allowing the formation of locally commensurate contact regions through in-plane strain fields~\cite{Guerra2017b}. In this respect, it has been shown that structural lubricity effects are enhanced in misaligned configuration owing to the reduced size of the moir\'e pattern and effective mutual cancellation of lateral forces~\cite{Mandelli2017,Ouyang2018}.
Besides, our reported static friction force values indicate that faceted MWNTs experience larger barriers against motion than collapsed DWNTs. Moreover, as previously observed in the superlubric case of graphitic nanoribbons~\cite{Gigli2017}, here we show that also in the case of rolling motion static friction oscillates with the NT contact width with a periodicity precisely dictated by the moir\'e characteristic length.

Kinetic friction can be interpreted from our results as sum of the separate contributions from sliding and rolling motions, which have intimately different origins. In a purely sliding motion, dissipation originates at the NT interface due to the substrate corrugation potential. This kind of contribution is roughly constant in the size range considered due to superlubricity effects.
Conversely, in a purely belt-like rolling motion the interface atoms are at rest, while all the dissipation originates from NT internal deformations. This contribution is proportional to the NT size and to the number of NT walls, so producing a linear growth of $F_{\rm kinetic}$ with the NT size in DWNTs, and an unprecedented supra-linear growth in the aligned MWNTs.
In the resulting picture, kinetic friction results lower in large misaligned (sliding) and in small aligned (rolling) MWNTs, with a threshold size of 7-8 walls connecting the two.

Some evidence to the above results, we note, may be obtainable in AFM nanomanipulation experiments~\cite{Vanossi2018}. Rolling motion might conceivably be addressed by labelling a NT periphery, or indirectly by QCM experiments~\cite{Krim1988}. Extending the initial tip-based studies~\cite{Falvo1999}, the possibility to laterally push NTs deposited at different orientation with the substrate seems entirely open to experimental verification. 
In real -- non-periodic -- systems, the aligned geometry is expected to be energetically more favourable for armchair NTs. More generally, the optimal orientation will depend on the chirality of the outermost wall~\cite{Chen2013} while additional metastable orientations will appear, similarly to the case of physisorbed nanoribbons~\cite{Gigli2017}.
Note that during nanomanipulation experiments in which long NTs are pushed from one edge by a sharp tip, the resulting motion can generally include sliding, rolling, and misalignment, exploring a multitude of interface configurations between the two ideal cases considered here. The results in our work can therefore serve as reference in such case of generic NT motion.

Roughness in graphene could also influence our ideal picture. For example, buckling effects are known to emerge especially in the case of applied vertical loads~\cite{Lee2010}. Also, rippling effects related to moir\'e have been observed in BN/graphene interfaces, even though these being limited to roughness heights of only $\sim$40\,pm~\cite{Li2015}, with expected small effects on the frictional response. Finally, adhesion effects could perturb the relaxed configuration by bending the substrate in order to maximise adhesion. Preliminary tests in this direction show that such perturbation is limited to a small region at the edges of the interface contact, indicating that such bending would influence in particular small size NTs, while larger NTs -- closer to typical experimental sizes -- would only be affected in a minor way. All these perturbative effects could be subject of future research.


\bibliographystyle{unsrt}
\bibliography{biblio_clean.bib}

\end{document}